# A Review and Analysis of Eye-Gaze Estimation Systems, Algorithms and Performance Evaluation Methods in Consumer Platforms

Anuradha Kar, *Student Member, IEEE,* Peter Corcoran *Fellow, IEEE*

*Abstract*— **In this paper a review is presented of the research on eye gaze estimation techniques and applications, that has progressed in diverse ways over the past two decades. Several generic eye gaze use-cases are identified: desktop, TV, head-mounted, automotive and handheld devices. Analysis of the literature leads to the identification of several platform specific factors that influence gaze tracking accuracy. A key outcome from this review is the realization of a need to develop standardized methodologies for performance evaluation of gaze tracking systems and achieve consistency in their specification and comparative evaluation. To address this need, the concept of a methodological framework for practical evaluation of different gaze tracking systems is proposed.**

## I. Introduction

ADVANCES in eye gaze tracking technology over the past few decades have led to the development of promising gaze estimation techniques and applications for human computer interaction. Historically, research on gaze tracking dates back to the early 1900s, starting with invasive eye tracking techniques. These included electro-occulography using pairs of electrodes placed around the eyes or the scleral search methods that include coils embedded into a contact lens adhering to the eyes. The first video based eye tracking study was made on pilots operating airplane controls in the 1940s [1]. Research on head-mounted eye trackers advanced in the 1960s and gaze tracking developed further in the 1970s with focus on improving accuracy and reducing the constraints on users. With increasing computing power in devices, real time operation of eye trackers became possible during the 1980s. However till this time, owing to limited availability of computers, eye tracking was mainly limited to psychological and cognitive studies and medical research. The application focus towards general purpose human computer interaction was sparse. This changed in the 1990s as eye gaze found applications in computer input and control [2]. Post 2000, rapid advancements in computing speed, digital video processing and low cost hardware brought gaze tracking equipment closer to users, with applications in gaming, virtual reality and web-advertisements [3].

Eye gaze information is used in a variety of user platforms. The main use cases may be broadly classified into (i) desktop computers [4]–[6], (ii) TV panels [7][8], (iii) head mounted [9]–[12] (iv) automotive setups [13]–[17] (v) handheld devices [18][19]. Applications based on desktop platforms involve using eye gaze for computer communication and text entry, computer control and entering gaze based passwords [20]. Remote eye tracking has recently been used on TV panels to achieve gaze controlled functions, for example selecting and navigating menus and switching channels. Head-mounted gaze tracking setups usually comprise of two or more cameras mounted on a support framework worn by the user. Such systems have been extensively employed in user attention and cognitive studies, psychoanalysis, occulo-motor measurements [2], virtual and augmented reality applications [21][22]. Real time gaze and eye state tracking on automotive platforms is used in driver support systems to evaluate driver vigilance and drowsiness levels. These use eye tracking setups mounted on a car's dashboard along with computing hardware running machine vision algorithms. In handheld devices such as smartphones or tablets, the front camera is used to track user gaze to activate functions such as locking/unlocking phones, interactive displays, dimming backlights or suspending sensors[18][23].

Within each of these use cases there exists a wide range of system configurations, operating conditions and varying quality of imaging and optical components. Furthermore, the variations in eye-movement and biological aspects of individuals lead to challenges in achieving consistent and repeatable performance from gaze tracking methods. Thus, despite several decades of development in eye gaze research, performance evaluation and comparison of different gaze estimation techniques across different platforms is a still a difficult task [24][25].

In order to provide insight into the current status of eye gaze research and outcomes, this paper presents a detailed literature review and analysis that considers algorithms, system

Submitted revised manuscript on 21st June 2017 to IEEE Access for review. Accepted on 24th July 2017.
The research work presented here was funded under the Strategic Partnership Program of Science Foundation Ireland (SFI) and co-funded by FotoNation Ltd. Project ID: 13/SPP/I2868 on "Next Generation Imaging for Smartphone and Embedded Platforms".
Anuradha Kar is with the Center for Cognitive, Connected, and Computational Imaging, Department of Electrical & Electronic Engineering, National University of Ireland, Galway. (email: a.kar2@nuigalway.ie)
Peter Corcoran is with the Center for Cognitive, Connected, and Computational Imaging, Department of Electrical & Electronic Engineering, National University of Ireland, Galway. (email: peter.corcoran@nuigalway.ie)

configuration, user conditions and performance issues for existing gaze tracking systems. Specifically, use-cases based on five different eye gaze platforms are considered.

The aim of this work is to gain a realistic overview of the diversity currently existing in this field and to identify the factors that affect the practical usability of gaze tracking systems. Further, this review highlights the need for developing standardized measurement protocols to enable evaluation and comparison of the performance and operational characteristics of different gaze tracking systems. In this paper, first the diversity and standardization issues in different aspects of eye gaze research are discussed and then the idea of a performance evaluation framework is proposed. This framework includes several planned and ongoing experiments that are aimed at practical evaluation of any gaze tracker. Our goal is to encourage further discussion and additional contributions from researchers in this field.

There have been detailed review works on eye gaze made in the last few years such as [3], [26]–[28] which discussed about recent developments in gaze tracking methods, comparing different estimation techniques, setups, applications and challenges involved in using gaze as an input modality. Hansen [26] provides an in-depth review on different eye models, eye detection techniques and models for gaze estimation, along with a summary of gaze applications. It also discusses inaccuracies in gaze tracking arising from the eye model components, jitter and refraction due to user wearing glasses. However, our work differs on several grounds from these reviews. Firstly, our review is specifically aimed towards highlighting the issues affecting realistic performance evaluation of gaze tracking systems, such as ambiguous accuracy metrics and un-accounted error sources. Secondly, we do a detailed classification of four different eye gaze research platforms. Extensive literature resources are collected and analyzed for each platform with the aim to understand the factors that affect the performance of a gaze based system in each of these. Thirdly, we present our survey in a statistical format that shows the lack of standardization in eye gaze research as a quantitative observation. Also our survey includes research works published until 2017 to make it exhaustive and up-to-date. Finally, our review not only provides an overview of the current status of eye gaze research but also forms the foundation of a performance evaluation framework for eye gaze systems which is proposed by us in Section VI of this work and is currently under development.

The paper is organized as follows: Section II presents a brief overview on eye movements, gaze tracking systems and accuracy measures used in contemporary gaze research. In Sections III and IV, several gaze tracking algorithms are categorized and key research works on the implementation of gaze tracking in five different user platforms are reviewed. In Section V, the factors limiting practical performance of gaze tracking in different user platforms are analyzed and issues with diversity in gaze accuracy metrics are discussed. The background and concept of a methodological framework for practical evaluation of eye gaze systems is presented in Section VI. We note here that the scope of this review excludes gaze tracking for clinical and neurological directions, retinal imaging and studies on children and patients.

## II. Eye Gaze Tracking Fundamentals

### A. Types of eye movements studied

Several types of eye movements are studied in eye gaze research and applications to collect information about user intent, cognitive processes, behavior and attention analysis [28]–[31]. These are broadly classified as follows: 1. Fixations: These are phases when the eyes are stationary between movements and visual input occurs. Fixation related measurement variables include total fixation duration, mean fixation duration, fixation spatial density, number of areas fixated, fixation sequences and fixation rate. 2. Saccades: These are rapid and involuntary eye movements that occur between fixations. Measurable saccade related parameters include saccade number, amplitude and fixation-saccade ratio 3. Scanpath: This includes a series of short fixations and saccades alternating before the eyes reach a target location on the screen. Movement measures derived from scanpath include scanpath direction, duration, length and area covered 4. Gaze duration: It refers to the sum of all fixations made in an area of interest before the eyes leave that area and also the proportion of time spent in each area. 5. Pupil size and blink: Pupil size and blink rate are measures used to study cognitive workload. Table I presents the characteristics of different eye movements and their applications.

### B. Basic setup and method used for eye gaze estimation

Video based eye gaze tracking systems comprise fundamentally of one or more digital cameras, near infra-red (NIR) LEDs and a computer with screen displaying a user interface where the user gaze is tracked. A typical eye gaze tracking setup is shown in Fig. 1. The steps commonly involved in passive video based eye tracking include user calibration, capturing video frames of the face and eye regions of user, eye detection and mapping with gaze coordinates on screen. The common methodology (called Pupil Center Corneal Reflection or PCCR method) involves using NIR LEDs to produce glints on the eye cornea surface and then capturing images/videos of the eye region [26][32]. Gaze is estimated from the relative movement between the pupil center and glint positions. External NIR illumination with single/multiple LEDs (wavelengths typically in the range 850+/- 30 nm with some works such as [33] using 940 nm) is often used to achieve better contrast and avoid effects due to variations induced by natural light. Webcams are mostly used; those operate at 30/60 fps frame rate and have infrared (IR) transmission filters to block out the visible light. Different gaze tracking methods are discussed in detail in Section III.

The user-interface for gaze tracking can be active or passive, single or multimodal [34]–[36] . In an active user interface, the user's gaze can be tracked to activate a function and gaze information can be used as an input modality. A passive interface is a non-command interface where eye gaze data is collected to understand user interest or attention. Single modal gaze tracking interfaces use gaze as the only input variable whereas a multimodal interface combines gaze input along with mouse, keyboard, touch, or blink inputs for command.

TABLE I
CLASSIFICATION OF EYE MOVEMENTS

| Eye movement type | Movement rate | Latency/duration of occurrence | Functionality/ Significance | Applications in Human Computer interaction |
|---|---|---|---|---|
| Fixation | < 15 – 100 deg/ms | 180- 275 ms | Acquiring information, Cognitive processing, attention | Browsing information, reading, scene perception |
| Saccade | 100- 700 deg/sec | Latency-200 ms, duration: 20–200 ms | Moving between targets | Visual search |
| Smooth pursuit | < 100 deg/sec based on target speed | 100 ms | Following moving targets | Gaze based drawing, steering |
| Scanpath | -- | -- | Scanning, direct search | Assessing user behavior, user interface and layout quality |
| Gaze duration | -- | -- | Cognitive processing, conveying intent | Item selection, text/number entry |
| Blink | 12 -15 per min | 300 ms | Indicates behavioral states, stress | Eye liveliness detection, activate command/control |
| Pupil size change | 4-7 mm/sec | 140 ms | Cognitive effort, representing micro-emotions | Assessing cognitive workload, user fatigue, command/control |

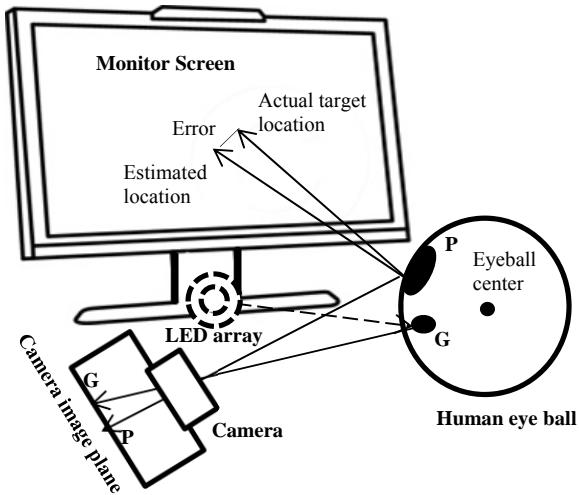

Fig.1. Schematic diagram of a typical gaze tracking system. P is the pupil of the human eye ball and G is the glint location formed on the cornea, which are imaged on the camera plane. The figure also shows the error in gaze estimation as the deviation between actual & estimated gaze locations

### C. Calibration

A generalized structure and model of a human eye is shown in Fig. 4a. The eye parameters typically required in gaze estimation are pupil center, center of curvature of cornea, the optical and the visual axes [32]. The posterior of the eyeball is called retina and the center of the retina with highest visual sensitivity is called the Fovea. The line joining the fovea with the center of corneal curvature is called the visual axis. Optical axis is the line passing through the pupil center and center of corneal curvature as shown in Fig 4a. The visual axis determines the direction of gaze and deviates from the optical axis. This offset is known as the kappa angle and measures around 5 degrees [37] but is dependent on each user. In gaze estimation, the pan and tilt components of the kappa angle are unique to each user and the visual axis cannot be estimated directly. The visual axis and the kappa angle therefore have to be obtained for each person though a process called calibration which has to be done at the start of an eye tracking procedure. Calibration is performed by showing the user a set of specific targets distributed over the front screen (as shown in Fig. 2) and the user is asked to gaze at them for a certain amount of time [38]. The tracker camera captures the various eye positions for each target point which are then mapped to the corresponding gaze coordinates and thus the tracker learns this mapping function. Calibration routines differ in the number and layout of target points, user fixation duration at each point and type of mapping algorithm used.

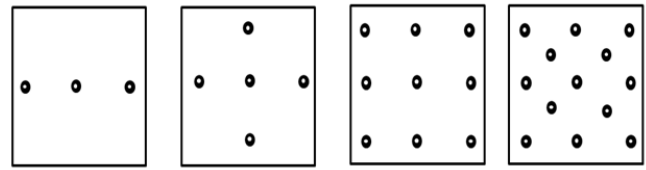

Fig.2. Calibration screen with 3, 5, 9, 13 target points

### D. Correspondence of eye gaze with head positions

The gaze location of a user depends both on the gaze direction and also on the head orientation[39]. In methods which use PCCR techniques, if the user moves their head with respect to the tracker-camera axis while looking at the same point on the front screen, the glint vectors with respect to the pupil centers (for two different eye locations produced by head movement) will be different from each other. Therefore the estimated gaze locations will be inaccurate. Eqn 1 presents the relationship between user reference gaze directions ($d_{kref}$), head pose direction $d_k$ and actual gaze direction ($d_{kgaze}$) which is a result of both head and eye rotation, as shown in Fig. 3. The effect of head movement has to be compensated before applying the gaze mapping algorithm or a chin rest for fixing head pose has to be used.

$$d_k - d_{kref} = k(d_{kgaze} - d_{kref}) \qquad (1)$$

where k is a parameter related to head pan and tilt reported in [39] with values 0.5 and 0.4 respectively.

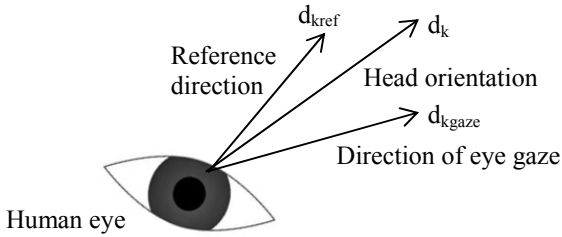

Fig.3. Relation between gaze direction & head pose

### E. Estimation of gaze tracking accuracy

In a typical eye gaze tracking operation, a user gazes at an interface on a computer screen which provides them with visual stimulus in the form of a set of targets or a scene. Gaze tracking accuracy is estimated as the average difference between the real stimuli positions and the measured gaze positions, which also provides an idea about the performance of the system.

In conventional literature gaze tracking accuracy measures are presented in different ways e.g. angular accuracy in degrees, distance accuracy in cm or distances in pixels. These accuracy estimate calculations are shown below. In practice, calculations are made separately for both eyes. For brevity, single calculations are presented and the same equation holds for both right and left eye. $POG.X_{left}$, $POG.Y_{left}$, $POG.X_{right}$, $POG.Y_{right}$ are the measured X,Y coordinates of the left and right eye's point of gaze (PoG). The mean gaze coordinates considering both eyes are $POG.X$ and $POG.Y$. $dist$ is the distance of the eye from the screen and $mean\_dist$ is the mean distance of eye from the tracker. The x/y pixels are the pixel shifts in x/y directions and offset is the distance between the tracker sensor and lower edge of display screen. Further details on these calculations can be found in [40].

**Gaze point coordinates:**

$$POG.X = mean\left(\frac{POG.X_{left} + POG.X_{right}}{2}\right) \quad (2)$$

$$POG.Y = mean\left(\frac{POG.Y_{left} + POG.Y_{right}}{2}\right) \quad (3)$$

**Pixel accuracy (Pix_acc):**

$$Pix\_acc = \sqrt{\left((target.X - POG.X)^2 + (target.Y - POG.Y)^2\right)} \quad (4)$$

**On Screen Distance (OSD):**

$$OSD = pixelsize\sqrt{\left(\left(POG.X - \frac{x_{pixels}}{2}\right)^2 + \left(y_{pixels} - POG.Y + \frac{offset}{pixelsize}\right)^2\right)} \quad (5)$$

**Angular accuracy (Ang_acc):**

$$\text{Gaze angle } (\theta) = \tan^{-1}(OSD/dist) \quad (6)$$

$$Ang\_acc = (pixelsize * Pix\_acc * \cos(mean(\theta))^2)/mean\_dist \quad (7)$$

## III. EYE GAZE ESTIMATION ALGORITHMS

Eye gaze tracking algorithms comprise of corneal reflection based methods which use NIR illumination to estimate the gaze direction or the point of gaze using polynomial functions, or a geometrical model of the human eye. 2D regression, 3D model, and Cross ratio based methods fall into this category. Another class of methods uses visible light and content information (e.g. local features, shape, texture of eye regions) to estimate gaze direction, e.g. appearance and shape based methods. The five different gaze tracking methods have their own advantages and disadvantages which are briefly discussed at the end of this section and summary of some key works on the development of these algorithms are presented in Table II.

### A. 2D regression based methods

In regression based methods, the vector between pupil center and corneal glint is mapped to corresponding gaze coordinates on the frontal screen using a polynomial transformation function. This mapping function can be stated as: $f: (X_e, Y_e) \rightarrow (X_s, Y_s)$ where $X_e, Y_e$ & $X_s, Y_s$ are equipment and screen coordinates respectively. The relation can be presented as described in [41] and [42]:

$$X_s = a_0 + \sum_{p=1}^{n} * \sum_{i=0}^{p} a_{(i,p)} X_e^{p-i} Y_e^i \quad (8)$$

$$Y_s = b_0 + \sum_{p=1}^{n} * \sum_{i=0}^{p} b_{(i,p)} X_e^{p-i} Y_e^i \quad (9)$$

Where n represents polynomial order, $a_i$ & $b_i$ are the coefficients. The polynomial is optimized through calibration in which a user is asked to gaze at certain fixed points on the frontal screen. The order and coefficients are then chosen to minimize mean squared difference ($\varepsilon$) between the estimated and actual screen coordinates (with known camera coordinates of user gaze), which is stated as:

$$\varepsilon = (X_s - Ma)^T(X_s - Ma) + (Y_s - Mb)^T(Y_s - Mb) \quad (10)$$

Where a and b are the coefficient vectors and M is the transformation matrix given by:

$$a^T = [a_0\ a_1 \ldots a_m],\ b^T = [b_0\ b_1 \ldots b_m] \quad (11)$$

$$M = \begin{bmatrix} 1 & X_{e1} & Y_{e1} & \cdots & X_{e1}^n & \cdots & X_{e1}^{n-i}Y_{e1}^i & \cdots & Y_{e1}^n \\ 1 & X_{e2} & Y_{e2} & \cdots & X_{e2}^n & \cdots & X_{e2}^{n-i}Y_{e2}^i & \cdots & Y_{e2}^n \\ \vdots & \vdots & \vdots & \cdots & \vdots & \cdots & \cdots & \cdots & \cdots \\ 1 & X_{eL} & Y_{eL} & \cdots & X_{eL}^n & \cdots & X_{eL}^{n-i}Y_{eL}^i & \cdots & Y_{eL}^n \end{bmatrix} \quad (12)$$

Where M is the transformation matrix, m the number of coefficients and L the number of calibration points [42]. The coefficients can be obtained by inverting the matrix M as [43]:

$$A = M^{-1}X_s,\ b = M^{-1}Y_s \quad (13)$$

Cherif et al. [42] used a 5x5 point calibration routine and 2 higher order polynomial transformations while Villanueva [44] studied effects of head movement on system accuracy with a 4x4 and 8x8 grid. Blignaut [41] also compared several mapping functions and calibration configurations with a 15x9 point grid and determined that number and arrangement of calibration targets and components of the mapping function play very important roles in determining overall accuracy of tracker. Robust and accurate gaze estimation under head movement was also achieved using neural networks by Ji [45] and Chuang [46].

Some other key works in this class of gaze estimation methods include Ma et al. [47] which introduces a 2D mapping algorithm that can handle unconstrained head movements and distorted corneal reflections due to various noise effects. It uses multiple geometrical transformation based mapping of CRs and demonstrates high reliability measures for different user distances and loss of CRs due to head/eye motion. A calibration free algorithm is detailed in Bennett [48] using SVMs which is robust to natural head movement and achieves high accuracy (1.5 degrees) in presence of head movement Villanueva[44] provided an exhaustive and detailed review of mapping equations and their impact on gaze tracking system response. The paper reports 400000 calibration functions, mapping orders and features to compare their impact on accuracy measures. Another important work is by Zhu et al. [49] which presents a very high resolution gaze estimation method resistant to head pose without requiring geometrical models. In [50] an improved three layer artificial neural network is used to estimate the mapping function between gaze coordinates and the pupil-glint vector. This method is shown to achieve better accuracy than simple regression based methods. A new approach involving only 2 light sources instead of four is implemented in [51] where two IR LED induced glints are real and the other two are virtual ones computed mathematically. This is done to make the algorithm more suitable for consumer applications and simplify hardware. The PoG is then estimated using mapping function to relate the four glint locations and screen through a calibration process. A modified PCCR method is implemented in [52] to have improved tracking accuracy and suitable for indoor and outdoor use. In this, adaptive exposure control is proposed since the IR LED brightness variations within a PCCR based setup affects pupil detection and gaze tracking to a large extent.

### B. 3D model based methods

These methods use a geometrical model of the human eye to estimate the center of the cornea, optical and visual axes of the eye (Fig. 4a) and estimate the gaze coordinates as points of intersection where the visual axes meets the scene. 3D model based methods can be categorized on the basis of whether they use single or multiple cameras and type of user calibration required.

3D model based methods using single camera have been reported by Martinez [53], Guestrin[32] and Hennessey [54]. Single camera systems have simple system geometry, no moving parts and fast re-acquisition capabilities. For 3D gaze estimation in Martinez[53] a single camera and LED are used to achieve an accuracy of 0.5 degrees with user calibration. Guestrin presents a mathematical model to reconstruct the optical and visual axes of the user's eyes from the centres of the pupil and glint in the captured video frames and configuration of a remote gaze tracking system. The model considers single and multiple cameras and light sources in estimation of the point of gaze. It then demonstrates the gaze tracking performance of a system implemented using two NIR light sources and one camera using the model. Their method achieves an accuracy of around 0.9 degrees. The system proposed by Hennessey includes a single camera and multiple LEDs to achieve 3D gaze tracking with free head motion.

Multi camera methods achieve high accuracy and robustness against head movement but require elaborate system calibration procedures including calibration of cameras for 3D measurements, estimating positioning of LEDs and determining the geometric properties of the monitors and their relation with the cameras. Some key works using two or more cameras include Shih [55], Ohno[56], Beymer [57] and Zhu [37]. Ohno describes 3D gaze tracking allowing free head motion using simple two point calibration and a two camera system comprising of an eye positioning unit and a gaze detection unit. The eye positioning unit uses narrow field stereo cameras and controls direction of the gaze positioning unit to achieve head motion independent tracking.

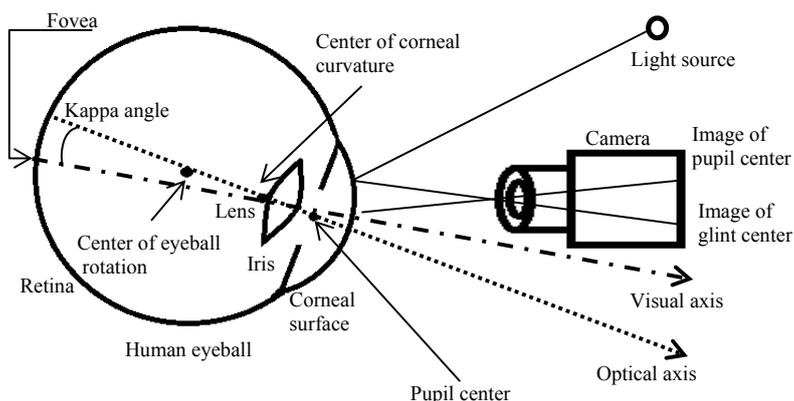

Fig. 4a

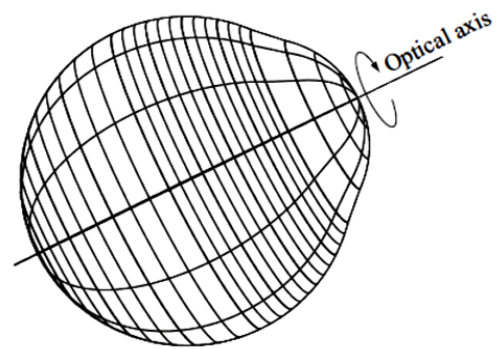

Fig. 4b

Fig.4a. Model of a human eye ball, eye parameters and setup elements used in 3D eye gaze tracking [32][56]. The optical axis is shown as the line joining the center of curvature of the cornea with the pupil center. The visual axis passes through the fovea and the center of corneal curvature. Kappa angle is the angular deviation between the optical and visual axis.

Fig.4b. An aspherical model of the cornea, as a surface of revolution about the optical axis of the eye [59]

A head pose free gaze tracking system is also implemented by Beymer with a wide angle stereo system for eye positioning and narrow angle stereo system for gaze detection. Pan and tilt directions of the narrow angle camera are controlled using rotating mirrors with galvo-motors. Shih and Liu [58] used a simplified eye model by Le Grand with two cameras and two LEDs to estimate the optical axis of the eye through solving linear equations. It uses single point calibration. The method in Zhu uses a gaze mapping function along with a dynamic head compensation model to update the gaze mapping function whenever the head moves to achieve tracking under natural head movement. It uses a 2 camera system with one time user calibration. With respect to the model of human eye, [32] provided evidence that a fully spherical corneal model will result in no impact of head movements on gaze estimation. They assumed an ellipsoidal model of the cornea and reported that gaze estimation errors increase with corneal asphericity and this also results in sensitivity of gaze estimation to head movements. An aspherical model of the cornea [59] is shown in Fig 4b and it is a surface of revolution about the optical axis of the eye. Use of this model showed to result in better accuracy, especially near the display corners as compared to a traditional 3D model based method.

Calibration free gaze estimation techniques have been proposed by Nagamatsu[60], Model[61] and Morimoto[4]. In [60], a calibration-free method is proposed using two pairs of stereo cameras, light sources and a spherical model of the cornea. One pair of cameras and two light sources are used for each eye to estimate the eye optical axis and the position of the center of the cornea. Optical axes of both eyes are measured using a binocular 3D eye model to estimate the point of gaze, achieving an accuracy of around 2.0°. Model & Eizenman also proposed a multiple camera based system to capture stereo images of the eye with corneal reflections. From the stereo eye images, eye features, such as the center of the pupil and corneal reflections are used to estimate subject-specific eye parameters. These parameters are then used with the eye features to estimate PoG. Their method is calibration free, has a tracking range of 3 to 5 meters and accuracy of less than 2 degrees. Morimoto proposed a method using two light sources and one camera that doesn't require user calibration for every session. It uses the Gullstrand model of the eye and ray tracing techniques to estimate the cornea and pupil centers. It achieves an accuracy of 2-4 degrees of visual angle dependent on the position of the light sources.

A new class of 3D gaze tracking has recently emerged with the usage of depth sensors in several works. These sensors comprise of an RGB camera and an infra-red depth camera. Typically resolution for the RGB camera is 640 x 480 pixels, with 45 degrees vertical and 58 degrees horizontal field of view. The depth camera resolution is about 1.5 mm at 50 cm. Gaze tracking using the consumer grade depth sensor (Kinect) is proposed in [62]. The method uses an eye model; 3D coordinates of eye features are obtained from Kinect and eye parameters like eyeball and pupil center are derived from a user calibration process. With this, 3D gaze coordinates are tracked in real time with a simple setup. Another work [63] reports the use of Kinect and a simple low cost setup for 3D model based gaze estimation allowing free head motion. It derives the 3D model parameters using convolution based means of gradients iris center localization method and uses a geometric constraints-based method to estimate the eyeball center. They assume that iris center points are distributed on a sphere originated from the eyeball center and the sizes of two eyeballs of a subject are identical. Kinect data is used to obtain 3D positions of person's head pose, iris and eyeball centers. [64] also uses a Kinect sensor and a model to estimate eyeball center by making users look at a target in 3D space. Kinect is used to build a head model to determine the eyeball center, detect the pupil center and determine 3D eye gaze coordinates in conjunction with the eye model.

### C. Cross-ratio based methods

These methods work by projecting a known rectangular pattern of NIR lights on the eye of the user and estimating the gaze position using invariant property of projective geometry. Four LEDs on four corners of a computer screen are used to produce glints on the surface of the cornea (Fig. 5). From the glint positions, the pupil and the size of the monitor screen, gaze location is estimated using two perspective projections. The first projection comprises of the virtual images of the corneal reflections of the LEDs (scene plane). The second projection is the camera projection, that is the images of the corneal reflections on the camera's imaging plane. With these two projections a single projective transformation relating the scene and camera image plane is obtained. Then the projection of the PoG on the scene plane to the image of the pupil center on the camera plane can be estimated[65].

Key works on the development and experimental verification of the cross ratio based methods can be found in Yoo [66] and Hansen[6]. Coutinho et al. presents a detailed analysis of methods and comparison of their accuracy in [67] [68]. They also suggest improvements by including a fifth LED on the optical axis of the camera and using a calibration procedure to improve accuracy[68]. Error compensation with polynomial-based regression have been proposed by Cerrolaza [44] or Gaussian process regression [6]. Error correction by homography mapping that eliminates the need of the fifth light source has been proposed by Kang [69].

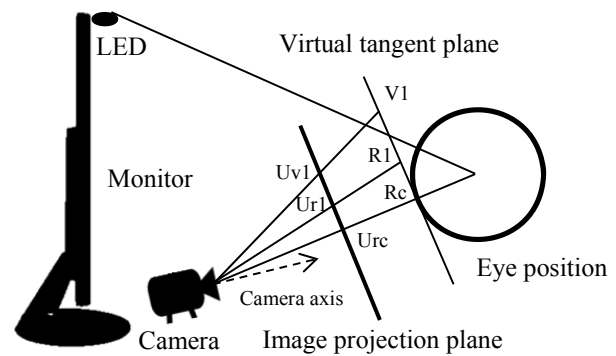

Fig.5. Setup for implementing cross ratio based gaze tracking [67]. Four light sources are used at four corners of the monitor screen (only one is shown here. V1 and R1 are the virtual projection and corneal reflection of L1, Rc is the reflection and the virtual projection of the LED fixed at the camera's optical axis. V1, R1 and Rc are projected to the image plane as Uv1, Ur1, and Urc.

### D. Appearance based methods

In appearance based methods the information from the eye region is represented using a model trained with a set of features extracted from eye images. In Bacivarov [70], a statistical model is used to represent shape and texture variations and trained using images of the eye region annotated with landmark points (Fig.6). The shape vector is the concatenated coordinates of all landmark points, stated as

$$s = (x_1, x_2, \ldots, x_L, y_1, y_2, \ldots, y_L)^T \qquad (14)$$

where L is the number of landmark points. The shape model is obtained by applying Principal Component Analysis (PCA) on the set of aligned shapes (equations derived from [70]):

$$s = \bar{S} + \varphi_s b_s, \qquad (15)$$

$$\bar{S} = 1/N_s \sum_{i=1}^{N_s} s_i$$

where $\bar{S}$ is the mean shape vector, and $N_s$ is the number of shape observations; $\varphi_s$ is the matrix having the eigenvectors as its columns; $b_s$ is the set of shape parameters. Similarly, the texture vector defined for each training image is: $t = (t_1, t_2, \ldots, t_p)^T$ (p : number of texture samples). The texture model is derived by means of PCA on the texture vectors as ($N_t$: number of texture observations, , $\bar{T}$ : mean texture vector)

$$t = \bar{T} + \varphi_t b_t \qquad (16)$$

$$\bar{T} = 1/N_t \sum_{i=1}^{N_t} t_i$$

The sets of shape and texture parameters ($b_t$) describe the appearance variability of the model:

$$c = \begin{pmatrix} W_s b_s \\ b_t \end{pmatrix} \qquad (17)$$

($W_s$ is the vector of weights). This is the statistical model that an Active Appearance Model (AAM) algorithm uses to best fit the model to a new eye image.

An active appearance based method for retrieving eye gaze from low resolution videos is presented in [71]. Global and local appearance models are trained and fitted for the whole face as well for capturing variance of the face and eye regions. For classifying the eye gaze into six directions, two different approaches are adopted. Gaussian Mixture Models (GMMs) are trained for large changes in gaze angles and for small gaze changes a Histograms of Oriented Gradients based method are tested. A method for 3D gaze tracking without use of active illumination is proposed in [72]. In this a synthetic iris appearance fitting method is introduced that computes the 3D gaze direction from iris shape. The method synthesizes a set of iris appearances and then fits the best solution to the captured eye image. This is claimed to remove unreliable iris contour detection problems arising in simple ellipse fitting and requirement of high resolution images by other methods. Once the iris contour is accurately estimated, a 3D eyeball model is used to estimate gaze from the captured eye image using the iris center/shape information.

Several appearance based methods report use of local features with Support Vector Machine (SVM)s for classification of gaze direction. These include Tang [73] in which an AAM is used to locate the eye region using 36 feature points that represent the contour of eyes, iris size, iris location, and position of pupils. Gaze direction is estimated from 2D coordinates of feature points and is classified using an SVM. In Wang [74] Local-Binary-Pattern (LBP) is used to calculate the texture features and a dual camera system is used to detect the space coordinates of the eyes. These two sets of information are fed into an SVM to classify the gaze direction under natural head movement. A novel method based on Local Binary Pattern Histogram (LBPH), is used in [75]. LBPH and PCA are used to extract eye appearance features and several classification methods based on SVM, neural networks and k-Nearest Neighbor (k-NN)s are tested for accuracy on a collected dataset for gaze estimation. The LBPH with SVM yields best accuracy. Chen & Liu[76] reports the use of a special kind of discriminatory Haar features and efficient SVMs (eSVMs) for implementing a computationally efficient gaze tracking method. Haar cascade is also used in [77]for real time gaze tracking. Rectangular features of the eye region are calculated to extract eye and pupil regions in an image which are mapped with the gaze coordinates on screen.

Neural network based approaches are used in [78] for head-pose tolerant gaze tracking. Training data comprised of cropped eye images of a user gazing at a given point on a computer screen and corresponding coordinates of that point. Neural networks are used along with a skin color model to detect the face and eye regions in [79]. An improved artificial neural network optimized using the Particle Swarm Optimization approach is used for fast, high accurate and robust gaze estimation with low resolution eye images in [80]. Some methods include use of 3D face models as in Lai [81] in which head pose free gaze estimation is implemented using a such a face model with head and eye coordinate systems. Both eye appearance and head pose are considered as components of a high dimensional head pose and eye appearance (HPEA) space. A Neighbourhood Approximation Forests (NAF) approach is used to model the neighbour structure of the HPEA space followed by Adaptive Linear Regression to estimate gaze direction. Other approaches report use of a deformable model [82] and Genetic algorithms[83].

In recent times, deep learning (DL) and convolutional neural network (CNN) based methods have been proposed for gaze estimation. In [84], a three stage CNN model is used to classify seven gaze directions from images taken with low cost webcams without need for calibration. Gaze tracking for a near eye display robust to illumination, skin and eye color variations and occlusion is implemented using CNNs in [85]. The CNN is used to learn the mapping from eye images to gaze position and comprises of two convolutional layers and two pooling with a fully connected layer at the end.

In [86] a smartphone based app is created to collect eye images from 1450 participants which is used to train a CNN based gaze tracker that can run in real time and without calibration. The dataset comprises of images with widely varying backgrounds, lighting and head motion and the network is trained with crops of both eyes and the face region. In [87]an extensive eye gaze dataset is built and a multi-modal CNN based method is tested. The dataset contains more than 200,000 images with variable illumination levels and eye appearances. The CNN uses a convolutional layer followed by a max-pooling layer and second convolution layer followed by a max-pooling layer, finally with a fully connected layer. The CNN learns the mapping between input parameters, i.e., 2D head angle, eye image and gaze angle (output).

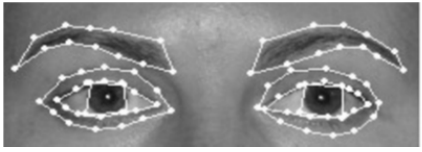

Fig.6. Image fitted with an Active Appearance model of the eye region [70]

*E. Shape based methods*

These methods employ deformable templates of the eye region, using two parabolas for eye contours (Fig.7) and circle for the iris, and fitting them to an eye image [88]–[91]. The procedure is to find the similarity between the template of a chosen region with images of that region.

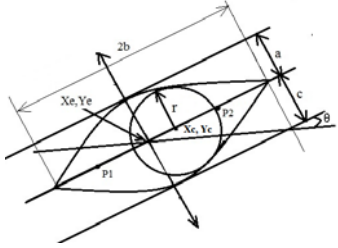

Fig.7. Template of an eye -region [88]. $X_c,Y_c$ & $X_e,Y_e$ represent the center of the pupil and the eye respectively. $P_1$ and $P_2$ are foci of two parabolic sections and a, b ,c and θ their parameters, r is the radius of the pupil.

This can be done by normalized cross-correlation, modified cross-correlation or by mean square error calculation [89]. If the template of pixel intensities of a region is represented by T(u,v) and I(i,j) represents that of the captured image then S(i,j) is the similarity measure between the template and image. If cross correlation is used as similarity measure, then S is given by:

$$S(i,j) = \frac{<T \times I_T> - <T><I_T>}{\sigma(T)\sigma(I_T)} \quad (18)$$

where < > is the average operator and < x > is the pixel-by-pixel product given by:

$$<T> = \frac{1}{n}\sum_{u,v} T(u,v)$$

$$<T \times I_T> = \frac{1}{n}\sum_{u,v} T(u,v)\, I(i+u,j+v) \quad (19)$$

σ is the standard deviation of the area being matched.

$$\sigma^2(T) = \frac{1}{n-1}\sum_{u,v}(T(u,v))^2 - <T>^2 \quad (20)$$

The mean squared error similarity measure is given by:

$$S(i,j) = \frac{1}{n}\sum_{u,v}(T(u,v) - I(i+u,j+v))^2 \quad (21)$$

*F. Summary and discussions*

The different gaze estimation algorithms presented above have distinct characteristics, advantages and disadvantages.

The 2D regression based methods utilize the features of the human eye, like eye geometry, pupil contours and corneal reflections and can be implemented using a single camera and a few NIR LEDs. However, these techniques are very vulnerable to head movements and require users to hold their head very still using a head rest, chin rest or bite bar.

3D model based methods have tolerance towards user head movement and most of them allow free head motion. However the hardware requirements for implementing 3D and stereo gaze tracking methods are high as they need several light sources or multiple cameras.

Cross ratio based methods have advantages, e.g. they do not need an eye model or hardware calibration and allow free head motion. But they are affected by problems such as increased error with distance of user and user dependent factors.

Appearance model-based algorithms are non-PCCR methods that use the shape and texture properties of the eyes and position of the pupils relative to the eye corners to estimate gaze. These methods have low hardware requirements which make them suitable for implementation on platforms without a high-resolution camera or additional light sources. The disadvantage is that their accuracy is mostly lower than PCCR based methods that degrade with head movements, variation in illumination levels and for robust performance they need large training image databases.

Shape based methods have been implemented for 2D gaze estimation with low-resolution webcam images achieving accuracy around 2°. However, downside of these methods include problems due to head pose variations and eye occlusions, adapting to largely variable eye shapes, computational complexity and issues with model initialization.

In Table II, some key research works on the above gaze tracking algorithms are presented. Column II presents the reference to the individual papers; Columns III to V presents various characteristics of the methods reported in them. The table highlights basic features and differences among different gaze tracking algorithms. 2D regression and appearance based methods have simple setups but typically offer accuracy values around 2 or 3 degrees. The accuracy of 2D regression based methods can be improved by increasing number of calibration points and using a chin rest to obtain fixed head position. On the other hand, 3D and cross ratio methods require more elaborate setups but offer much better accuracy (around 0.5 degrees) and most of them allow head movement.

## IV. USER PLATFORMS IMPLEMENTING GAZE TRACKING

In this section, the user platforms where eye gaze tracking has been implemented are described and classified.

*A. Desktop based systems*

Applications of eye gaze on desktop systems fall into several categories, such as computer communication, password entry and psychoanalysis. Sibert and Jacob [35] developed gaze based fast object selection as a substitute for mouse. A gaze based application called MAGIC (Manual And Gaze Input Cascaded) is presented by Morimoto [92] in which gaze based pointing is reported to have higher speed and accuracy than manual pointing. In Ghani [93], a Hough transform based pupil detection for gaze based control of a mouse pointer is proposed. In Villanueva [94] evaluation studies on the use of eye gaze in video gaming control for target acquisition and tracking are made. Gaze input had a similar performance to the mouse and joysticks for big targets.

Kasprowsky [95] and Kumar [96] reported the use of gaze to enter a password using gaze tracking. A series of user fixations on specified digits formed the password sequence. Methods robust to shoulder surfing problem were reported by Bulling [97] where a computational model of visual attention is used to increase security. Applications of using eye gaze patterns for identifying user tasks, mental workload and inferring context of events and user distraction were reported by Iqbal [98]and Doshi [99].

TABLE II
CLASSIFICATION OF GAZE ESTIMATION ALGORITHMS

| Method category | Paper reference | Setup (camera, LED) | Accuracy/ metrics (*deg= degree) | Tested for following operating conditions |
|---|---|---|---|---|
| **2D Regression** | [43] | Single steerable camera, multiple LEDs with 2 face camera in stereo | 0.8 deg | Allows limited head movement |
| | [42] | LED with phototransistor | 2.5 deg | No head movement |
| | [49] | 1 video camera | 1.4 deg, | No head movement |
| | [47] | 4 LEDs and webcam | 1.3 deg | Large head movement |
| | [45] | 2 IR LED rings and 1 camera | 5-8 deg | Significant head movement |
| | [48] | Two video cameras | 1.5 deg | Natural head movement |
| | [46] | Two loops of infrared light sources | 20 pixels | Head movement allowed |
| | [41] | Single camera, single infrared light source | 0.87 deg | No head movement |
| | [44] | Webcam with built in LEDs | Nearly 2.5 deg | Limited head movement |
| | [51] | 2 IR light sources, 1 webcam | 1.11 deg | User distance |
| **3D model** | [56] | Pan tilt camera –stereo with LED arrays | 1.0 deg | Free head motion |
| | [53] | 1 camera 2 LEDs | 1-3 deg | Full head pose compensated |
| | [4] | 1 camera 2 LEDs | 1-3 deg | Full head pose compensation |
| | [146] | 1 camera no LED | 1-2 deg | No head pose variation |
| | [57] | 4 camera 2 LEDs, mirrors | 0.6 deg | Head pose compensated |
| | [58] | 2 camera 2 LEDs | Less than 1-2 deg | Head pose compensated |
| | [147] | 2 camera 2 LEDs | < 1-2 deg | Head pose compensated |
| | [148] | 3D face model , 2 camera, no LED | 1 deg | Head pose compensated |
| | [54] | 2 camera, 1 light, mirror | 3 deg | Head pose compensated |
| | [55] | 2 ring of lights, 2 cameras | 1.25 deg | Small head pose tolerance |
| | [63] | 1 depth sensor with integrated LEDs | 3.78 deg | Head movement, user distance |
| | [64] | 1 depth sensor with integrated LEDS | 5 deg | Head pose |
| **Appearance based** | [83] | Eye images | 4.87 % accuracy | ----- |
| | [149] | VGA camera | 3.5 deg | No head movement |
| | [145] | Webcam | 1- 2 deg | Slight head movement |
| | [5] | VGA camera | 4 deg | Allows head movement |
| | [78] | Eye images | 1.5 deg | Slight head movement |
| | [79] | Eye images | 2.3 deg | No head movement |
| | [150] | 1 camera, NIR illuminator | 0.5 deg | ---- |
| | [91] | Webcam | 2.5 | ---- |
| | [81] | Model, camera | 3.5 deg | Free head movement |
| | [151] | 2 cameras | < 3 deg | Free head movement |
| | [73] | Webcam | Recognition rate: 97 % | ---- |
| | [82] | 1 camera | Correct adaptation rate: 91% | Tolerant to head pose, occlusion, low image resolution, illumination change |
| | [76] | Image databases | > 96% | ----- |
| | [84] | Image databases | Recognition rate 97% | ----- |
| | [87] | Image databases | 6.3 deg | Head pose, illumination variations |
| | [71] | Image databases | < 7 deg | Head pose variations, occlusion |
| | [80] | Commercial camera | 94 % | ------ |

| Method category | Paper reference | Setup | Accuracy/ metrics (*deg= degree) | Tested for following operating conditions |
|---|---|---|---|---|
| **Cross Ratio based** | [65] | 4 IR LEDs, 1 camera | 0.3- 0.4 deg without / 1-2 with head movement | Head movement allowed |
| | [67] | Camera with 16 IR LEDs | 0.9 deg | Free head movement |
| | [68] | 1 camera, 7 LEDs | 1.4 deg | Head movement allowed |
| | [152] | Camera, 8 IR LEDs | 0.3-0.4 deg | Free head movement |
| | [6] | 1 camera, 4 light sources | 0.5 deg without / 3.5 deg with movement | Head movement allowed |
| | [66] | 2 cameras. 5 LEDs | 0.98 deg | Large head movement |
| | [69] | 1 camera, 4 NIR LEDs | 10.3 mm | None |
| **Shape based methods** | [88] | 1 camera | 85% | None |
| | [89] | Image frames | 4.5 pixels | None |

*B. TV and large display panels*

There are recent applications of long range gaze estimation that use corneal reflection (CR) techniques for tracking gaze on large displays and smart TVs. Gaze movements can be used to select and navigate menus, modify display properties, switch channels and understand user interests. In Park et al. [7]a robust pupil detection method for gaze tracking on large display is presented using a wide and a narrow view camera with Adaboost and CAMShift algorithms. In another work, Park [8] reports a system for gaze tracking on a large-screen 60 inch TV based on a 2D method with geometric transform, using pupil center and four corneal specular reflections.

*C. Head-mounted setups*

Head mounted gaze trackers are portable platforms with applications ranging from computer input, interactions in virtual environments, gaming controls, augmented reality and neuro/psychological research. The general setup includes two cameras; one (eye camera) pointed at the wearer's eye, to detect the pupil; and the other (scene camera) capturing the wearer's point of view, with sometimes additional components like NIR light sources and hot mirrors. Head-mounted gaze trackers have been implemented as attachment-free, mobile, low-cost, lightweight devices with simple hardware and software. Also they are known to provide high accuracy gaze information in unconstrained settings.

3D gaze estimation with head-mounted trackers have been reported in several papers including [9], [100]–[104]. Algorithms for high accuracy and 3D eye tracking proposed by Park [100] are based on 3D human eye model and Purkinje images. Pupil size and Purkinje images are fed as inputs to a multi-layer perceptron to estimate the depth location of gaze followed by 2D gaze coordinates. Takamatsu [104] describes a Visual SLAM technique to estimate the user head pose and determine 3D point of regard of the user. It also includes a 3D environment to detect objects of focus and visualization of a 3D attention map in a real environment. Lanata [9] presents a stereo-vision based method that implements a novel binocular system comprising of two mapping functions: linear and quadratic for depth estimation of gaze locations in 3D space.

The performance compared to other 3D methods show significant improvement in accuracy. In [103], 3D gaze tracking with multiple calibration planes is implemented with a single eye view camera and four IR markers. The system operates in monocular and binocular mode independently. A low-cost 3D eye tracking solution is provided by [101] in which images from two eye cameras are used with intensity thresholding and blob-contour detection to determine pupil center coordinates. The 2D gaze location is then obtained by polynomial mapping and 3D gaze is obtained from the meeting point of the gaze vectors of both eyes.

2D methods and applications with head-mounted eye tracking have been proposed in a several works. [10] reports an ultra-low cost 2 camera based gaze tracking system that is easy to assemble and aligns scene and eye videos by using synchronized flashing lights. A novel smooth-pursuit-based calibration methodology is proposed that works with pupil detection to track gaze. A new high speed binocular eye localization scheme using simple components was proposed by Kiderman [105] using two cameras and two hot mirrors. It shows speed advancement over contemporary slow systems that do offline processing. Other advanced methods are proposed in Schneider & Bartl [106] where a head mounted camera system tracks and follows the user's gaze direction for natural exploration of a visual scene by capturing the perspective of the mobile user. Applications of this system include documentation of medical processes, sports etc. In Virtual and Augmented reality (VR and AR respectively) research, gaze tracking and gaze based functions are used to make the user experiences more immersive, natural and user interactions fast and efficient in a VR/AR environment. Applications for Augmented Reality are described in Choi [107] where the device uses a scene camera and eye tracker to estimate user gaze and blink-state for interacting with the AR environment. A head mounted system that employs gaze tracking for immersive and realistic gaming experience is proposed by Park [108] where cursor aiming in the gaming environment is controlled by gaze tracking. A wearable gaze tracking device for determining effectiveness of text layout and line spacing for analyzing reading behavior is described in Dengel [109]. Parkhurst and Babcock [11] presents the development of a head mounted system called OpenEyes with open hardware and software tools for gaze tracking. Several types of eye and head gestures, such as saccades, smooth pursuit and nod-roll are studied in [110] as interaction methods in a head mounted VR device. A head mounted display coupled with an eye tracker is used to study which of these gestures result in better user experience in a VR environment.

A facial re-enactment method for gaze aware VR system is presented in [111]. It does real time facial motion capture of a user wearing an HMD along with monocular eye tracking. The purpose is to achieve real time photo-realistic rendering of face and eye appearances as users change facial expressions and gaze directions. Gaze is estimated from images taken using monocular camera and IR LEDs using a hierarchical classification method. In [112] eye tracking is used to develop an immersive 3D user interface for VR and implement several multimedia applications along with usability testing. Eye gaze pattern is extracted along with fixations to implement 3D virtual menu selection and eye based typing for mail composition using a virtual keyboard. A saliency based gaze localization approach using deep learning has been proposed in [113]. In this, image features and head movements are inputs to a convolutional neural network to estimate gaze coordinates in a VR system.

To make eye tracking in HMD systems flexible with respect to various users, HMD drifts and tracker camera adjustments, [114] presents a method for automatic calibration of the eye camera in the HMD. With this, the users need not maintain a fixed head pose and relative movements of HMD and eye camera can be tolerated for reliable eye tracking. [115] makes use of gaze, dwell time and half blink information for the purpose of hands-free object selection with an optical see through head mounted AR device. The multiple input parameters are used to avoid accidental/unintentional selection of items. Pupil motion is tracked to estimate a user's viewing point, with dwell time and half blink detection successively used for the object selection process. Kalman filtering is used minimize pupil jitter and drifts. The gaze tracking accuracy of the system is reported to improve with more calibration points with the best being 0.39 degrees.

An innovative AR application combining eye tracking with a smartwatch for implementing a wearable context aware messaging service is done in [116]. A head-mounted eye tracker estimates the gaze and together with data from a scene camera it is used to track where the user is looking in the real world. The smartwatch works as a message input and output device in conjunction with the tracking device to embed/display messages with the "augmented objects. An AR application for reading and document retrieval using a head-mounted eye tracker along with a see through display is shown in [109]. Eye tracking helps to estimate the section of the document the reader is focused into, in real time. Associated information on the specific part of the document is retrieved and displayed on the HMD. A calibration procedure for the HMD is also mentioned. In this, a user is presented with several dots in the HMD and he/she has to click the position of each dot in a calibration window. The homography between the scene image and the HMD is then estimated to map the gaze position on the scene to that on the HMD. The system achieves accuracy sufficient to distinguish between alternate lines in a document. A hardware and software framework for AR displays implementing eye tracking and gaze based interactions can be found in [117]. In this, a see through HMD is integrated with stereo eye tracking and configurable optics for various see through configurations. The system is designed to have on demand zoom and FOV expansion in a see through AR system. Eye related parameters like gaze, squint and blinks are tracked for activating different functions such as binary, sub-regional and gradual zoom and capturing snapshots of the AR view.

### D. Automotive

Visual features of the face and eye regions of an automobile driver provide cues about their degree of alertness, perception and vehicle control. Knowledge about driver cognitive state helps to predict if the driver intends to change lanes or is aware about obstacles and thereby avoid fatal accidents. Gaze related cues that indicate driver attentiveness include: blink rate, temporal gaze variation, speed of eyelid movements and degree of eye openness. Several works for driver assistance systems have been reported that are based on video based gaze tracking using a variety of classifiers.

SVM based gaze classification is common on automotive platforms and has been reported in Fujimura[118], Park[119] and Chuang[120]. A real time gaze tracking method robust to variable illumination levels and driver wearing glasses in an automotive environment is reported in [118]. It uses Kalman filtering and mean-shift method to track driver eyes based on their locations in a previous frame. SVM with linear, polynomial and Gaussian kernels are used for eye verification. Park [119] proposed a robust SVM based driver gaze zone estimation system that works during day and night conditions and is robust to driver wearing eyeglasses. A multiclass SVM constructed from multiple two class (binary) SVMs was constructed to classify 18 driver gaze zones using a large database of 18000 gaze feature vectors. Another SVM based gaze zone classifier that takes face parts, i.e., mouth, eyes, and nose locations is reported in Chuang[120]. A multiclass linear SVM is used which inputs the feature descriptor to output 8 gaze directions. Use of other type of classifiers is reported in Tawari[121] and Kwak[122]. In [121] a distributed camera setup is used for estimating gaze zones combined with head pose dynamics. A random forest classifier is used with static and dynamic features (head pose angles and time series statistics) to classify 8 gaze zones for the driver. In [122] Viola Jones method is used to detect the face followed by using linear discriminant analysis (LDA) to extract features from the eye region for classification. *k*-nearest neighbour method with Euclidean norm is used to classify the obtained features into seven gaze directions.

Gaze estimation in automotive using Purkinje images and PCCR methods are reported in Yang[17], Salvucci[123], Batista[14],Choi[124] and Ji[125]. Several papers report the application of gaze information along with other facial and visual parameters for derivation of driver psychological/cognitive state. Key works include Bergasa [126] which reports a method for driver vigilance estimation using fuzzy classifier taking six parameters: Percent eye closure (PERCLOS), eye closure duration, blink frequency, nodding frequency, face position, and fixed gaze. A dynamic Bayesian Neural network approach is used in Yang [127] for driver fatigue detection using gaze location along with face detection, eye positioning and iris tracking. Ji[125] combines features such as eyelid and head movement, gaze, and facial expressions with a Bayesian network to determine driver fatigue. Another important work that considers eye movements for deriving driver cognitive state is [16]. It takes into consideration eye movement features like fixations, saccades, and smooth pursuits and calculates 16 different eye movement features. These are then fed to an SVM, static and dynamic Bayesian networks to do their performance comparison for prediction of driver distraction state. [128] presents a low cost system to detect eyes-off-the road condition of a driver using facial feature tracking, 3D head pose and gaze estimation. A monocular camera is installed close to the steering wheel for tracking a driver's facial landmarks and accurate estimation of driver pose, location and gaze direction. Then, with 3D analysis of car/driver geometry the driver's eyes off the road condition is predicted in real time. In [129], driver gaze behaviour is studied to evaluate driving performance of a user when they had to interact with a portable navigation system while driving. Glance frequency and glance time were estimated to study impact of varying display sizes and positions of the navigation device while in use during driving. A portable driver assistance system involving driver drowsiness detection and eye gaze tracking is implemented using a Raspberry Pi and machine vision algorithms in [130]. A new multi-depth calibration approach is presented in [131]for obtaining 3D user PoG with stereo face cameras and monocular scene cameras for performing driver intent and actions prediction.

Several works study the dynamics between head pose and gaze behavior of drivers. In [132] the relationship between gaze and head pose are studied and regression models are developed to predict gaze location from the position and orientation of a driver's head. In [133]head and eye poses of 40 driving participants are estimated from monocular video and relative significance of head vs eye movements in gaze classification is studied. It is observed that driver behavior can be grouped into two cases, i.e., when the head moves a lot and gaze classification is mostly affected by head pose. The other case is when the head stays still and only the eyes move and thus accurate classification requires study of eye pose.

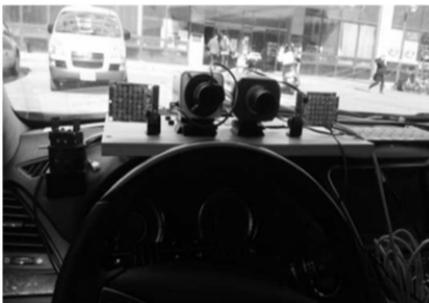

Fig.8. Driver monitoring cameras mounted on car dashboard [124]

*E. Hand-held devices*

Smartphones and tablets provide a unique paradigm for gaze tracking applications. Gaze tracking on handheld devices is done using the device front camera, one or more IR light sources and various computer vision algorithms. Key approaches for handheld-eye tracking are reported by Bulling [134] for tracking 3D gaze location from a video using a face and eye detector followed by edge detection and ellipse fitting to determine the eye limbus boundaries. In Lukander [135] a commercial eye tracker is integrated with a magnetic position tracking device to track positions of head and eyes. Komogortsev[136] uses a single perceptron neural network that maps the features of the eye region to a position on the screen. In Kikuchi [137] eye tracking on a tablet is realized with blob and contour detection for deriving iris contours. Then center and planar homography transform is used to convert the coordinate of iris center to display coordinates.

Several works use the Haar classifiers/Viola-Jones technique[138] such as Alimi [139] which describes a system for android tablets to track head motions and eye gaze gestures from video captured using device front camera. The system is claimed to be robust to user movement and lighting conditions. In Kavasidis[140] eye tracking for tablets is implemented using a Haar classifier based eye detection module in conjunction with the CAMSHIFT algorithm.

Potential applications of eye gaze in handheld devices are presented in Nagamatsu [19] describing a user-interface called Mobigaze that uses gaze to operate a handheld device. The system combines gaze and touch for operating the device interface thereby eliminating the Midas touch problem[31]. The Midas touch effect [31]refers to the phenomenon in which an eye-gaze-sensitive user interface cannot distinguish between user glances for collecting visual information and those for command input. Thus every user fixation may lead to activation even without the actual user intention. Another multimodal gaze based interaction method is presented by Drewes [141] for monitoring applications using dwell-time and gaze gesture. The EyePhone introduced by Miluzzo [142] controls the phone functions with only gaze and blinks and is free of user touch. It describes an application (Eyemenu) in which gaze direction is used to selectively access and highlight menu buttons. A gaze based user authentication scheme for smartphones is implemented in Wang [143] by making the user eyes to follow a moving target on the device screen. In Lai[144], a 3D image display is combined with gaze estimation to achieve adaptive 3D display on a mobile phone that shows different views to corresponding viewing angles. In Imabuchi[137] a gaze based communication interface is implemented for gaze based keyboard input and gaming.

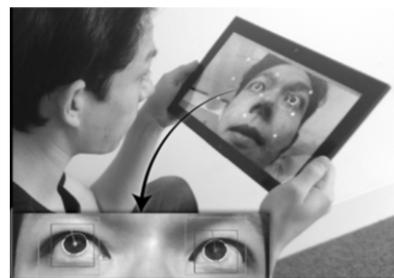

Fig.9. Eye tracking implemented in near real-time on a tablet. The tracking algorithm uses cascade classifiers and shape-based approaches to determine the eye region and centers. Elliptical model-fitting and 3D back-projections are then used to determine the eye optical axes and point-of-gaze [18].

Table III presents key information obtained from selected papers based on five gaze user platforms. It can be seen that different research works present their results in widely different formats and describe performance of their system under varying operating conditions.

TABLE III
SUMMARY OF RESEARCH ON EYE GAZE USE CASES IN VARIOUS PLATFORMS

| Use case/ Platform | Paper reference | Setup (camera, LED) | Accuracy/ metrics (*deg= degree) | Tested for following operating conditions |
|---|---|---|---|---|
| **Desktop** | [97] | 2 integrated infrared cameras | Euclidean distances, 21 and 46 pixels | Distance from tracker |
| | [95] | Commercial tracker, 1 camera | 61.1% | User dependent |
| | [96] | Commercial tracker, 1 camera | Error rate 15% | None |
| | [92] | Commercial tracker, 1 camera | Completion time, no. of hits/ misses | None |
| | [93] | 1 webcam | Recognition rate 100% | None |
| | [141] | Commercial eye-tracker, 1 camera | Percentage, time | None |
| | [94] | One camera | Mean error rate 22.5% | None |
| | [98] | Commercial tracker, 2 eye/1 scene cam | Percentage time spent | None |
| | [153] | 1 eye camera one scene camera | Number of fixations, time of fixations | None |
| | [99] | 2 cameras | Gaze shifts/ not mentioned | Head motion |
| **TV** | [8] | 1 camera, zoom-lens, 4 NIR illuminators | 60.4 pixel, 1.32 deg | None |
| | [7] | 1 wide & 1 narrow view camera | 5.75 pixels | Occlusion |
| **Headmounted** | [100] | Eye-capture camera and NIR (LED), zoom lens | 0.961 deg  1.601deg, (X, Y axis) | Distances from the screen |
| | [9] | Wireless camera, no LED | 0.85 deg | Head motion, light level changes |
| | [10] | 1 eye camera, 1 scene camera, 1 LED lamp | 0.752 deg | None |
| | [11] | Eye camera, scene camera, LED | 0.60, 1.03, 1.04 deg | Field of view |
| | [102] | 2 eye tracker cameras, 1 head camera, hot mirrors, 2 LED | 0.54 deg | Head motion, user motion |
| | [104] | 2 eye, 1 scene camera, 1 IR LED | 3.38 ± 2.38 deg | Head pose |
| | [105] | 2 camera, 2 hot mirrors, IR LED | less than 0.5 deg | Occlusion |
| | [107] | 1 eye camera, 1 scene camera | 17.6 pixels, 0.81 deg | Eye limits, midas touch |
| | [109] | 2 eye cameras, 1 scene camera. 6 IR LEDs | 0.5 deg | Distance from tracker, viewing angle |
| | [103] | 3 mini-cameras and 2 hot-mirrors | 1.2 deg | Distance from tracker |
| | [12] | 2 cameras | < 2 deg | None |
| | [154] | 1 eye camera, 1 scene camera, 1 LED | 0.66 deg | Distance from tracker |
| | [21] | 2 eye, 1 scene camera | 0.25 deg | Head pose |
| | [155] | 2 Cameras | 0.5 -3.5 deg | None |
| | [156] | Eye camera, scene camera, mirror | 0.66deg | None |

| Use case | Paper reference | Setup (camera, LED) | Accuracy/ Metrics (*deg= degree) | Tested for following operating conditions |
|---|---|---|---|---|
| **Automotive** | | | | |
| | [13] | 2 cameras | 5 deg | Head pose |
| | [17] | 2 cameras, 2 rings of LEDs | 98-100 % | None |
| | [157] | Commercial tracker -2+2 cameras | 3 deg | Head pose |
| | [124] | 1 Camera | Not stated | Head pose |
| | [158] | 2 cameras/ commercial tracker | 0.18 radians | None |
| | [159] | 1 camera | Not mentioned | Head motion |
| | [15] | 2 cameras- 1 facing driver, 1 looking out | Accuracy 94.9% | Head pose |
| | [127] | 1 camera | Detection ratio 85%. | Spectacles |
| | [125] | 1 camera, 2 rings of LEDs | Classification accuracy 97.5% | Head pose |
| | [126] | IR camera on dashboard, 2 IR LED rings | Correctness rate 90-100% | Head pose, glasses |
| | [119] | 2 NIR illuminators, a camera | Success detection rate 97% | Head pose/face pose |
| | [160] | 1 camera | Detection rate 86-100% | Head pose |
| | [14] | NIR LED rings,1camera | Not mentioned | Face pose |
| | [161] | 4 motion capture + 1 video camera | 10 pixels | None |
| | [16] | 2 cameras | 5 deg | Head pose |
| | [123] | Head-mounted eye tracker | 1 deg | None |
| | [162] | 1 camera | Not mentioned | Head pose, lighting |
| **Handheld devices** | [134] | Device front camera | 67.3% on a laptop | None |
| | [140] | Front camera of iPhone (1 camera) | 79.1 % accuracy | Hand motion |
| | [18] | 1 front camera | 6:88 deg | None |
| | [136]. | 1 tablet front camera | 3.47deg | None |
| | [137] | 1 front camera | 0.77 deg | None |
| | [144] | 1 device front camera | Average pixel error | Head positions |
| | [108] | 1 camera | Pixels | Head motion |
| | [142] | 1 camera | Percent accuracy 70-98% | Accuracy with screen positions |
| | [143] | 1 camera | True positive rate | None |
| | [19] | 2 cameras with IR-LED | Not discussed | |
| | [139] | 1 camera | Detection rate 60% | None |
| | [163] | 1 camera | Detection accuracy 80-100% | Distance from user |
| | [135] | Commercial tracker, 2 eye 1 scene camera | 0.15 Deg RMS | None |
| | [164] | 1 front camera | Recognition rate 93% | Distance |

## F. User platforms for gaze tracking: Summary of user configurations

The user conditions while they are using gaze tracking in the above five platforms are completely different. Similarly, the operating environment and tracking setup are also unique to each platform. Table IV presents typical system and user configurations for the five gaze tracking platforms including users' postures and viewing angles, screen sizes, typical distance between the user and the screen-camera setup. It can be seen that gaze tracking on different user platforms encounter wide range of operating conditions and therefore have diverse performance measures.

TABLE IV
FEATURES OF GAZE ESTIMATION SYSTEMS IN DIFFERENT USER PLATFORMS

| User platforms | Distance | Viewing angle (degree) | Screen size (inch) | User condition |
|---|---|---|---|---|
| **Desktop** | 30-50 cm | ~40 | 14-17 | Upright, sitting, static |
| **TV panels** | 2-5 m | 40-60 | 26-70 | Lean back, sitting, static |
| **Head mounted** | 2-5 cm | 55-75 | -- | Lean independent, sitting or standing, static or mobile |
| **Automotive** | 50 cm | 40-60 | -- | Upright, sitting, mobile |
| **Handheld devices** | 20-40 cm | 5-12 | 5-10 | Lean forward, sitting or standing, mobile |

Apart from the differences seen in Table IV above, Table III also shows the inconsistency in performance reporting formats among the different user platforms. Accuracies are reported in varied units –such as degrees, pixels, percentage of correct detection etc. Another feature as seen from Column 5 of this table is that only a few papers study the impact of operating conditions on the system performance. Further discussions about these aspects of gaze research are made in Section V.

Amongst the papers reporting in degree measures, a trend is seen in the accuracies for different eye gaze user platforms. Typically, it is seen that head mounted systems report better tracking accuracies of less than one degree amongst other platforms. For desktop systems it varies from 0.5 to 2 degrees of angular resolution and above 2 degrees for more dynamic platforms like automotive and handheld devices. Further, all platforms have unique setup requirements and users may assume various physical poses as seen in Table IV. Therefore, a general eye tracker may produce significantly different results depending on platforms.

## V. PERFORMANCE METRICS AND PLATFORM SPECIFIC ERROR SOURCES IN EYE GAZE RESEARCH

### A. Diversity of gaze estimation performance metrics in different user platforms

In contemporary literature, research works on gaze tracking present the results of their algorithm or application in several ways. The common measures used are angular resolution (in degrees), gaze recognition rates (in percentage) and minimum pixel shifts/distance between gaze and target locations. These metrics are no way correlated to each other and each research work defines some metrics, for example percent recognition rates or error rates in their own way. The consequence is that most gaze estimation methods cannot be inter-compared.

To better understand the diversity of performance measures used in gaze research, an information statistics is collected from our reviewed papers and presented in Table V. Firstly, the papers are grouped according to platforms and then the performance measures reported in each paper are classified according to the categories: degrees, percentage and "others" (pixel shifts/distance in mm). In Table V, Column 2 has the total number of research papers that was considered in the survey for each platform. Columns 3, 4 & 5 provide reference to papers that reported their performance measures in either "degrees", "percentage", or "other" formats respectively. Fig. 10 presents this statistics i.e. number of papers that reported accuracies in various formats for four different user platforms.

What can be observed from Table V is that there is no standard convention that is used to represent performance scores of eye gaze estimation algorithms in contemporary eye gaze research. For example, as seen in the table, out of total 69 papers from desktop/TV platform, 44 papers report tracking accuracy in degrees while 16 papers report them as gaze recognition rates and 9 papers in mixed units. Also this discrepancy is found in literature for all user platforms as seen from the rows 3 to 5 of the table and Fig. 10. From Fig. 10, we observe that degree measures are common units used for desktop and head-mounted platforms, whereas the representation is completely heterogeneous for automotive or handheld devices. The result of this inhomogeneity is that, the stated performance from a large volume of research in this field can neither be compared nor interpreted quantitatively.

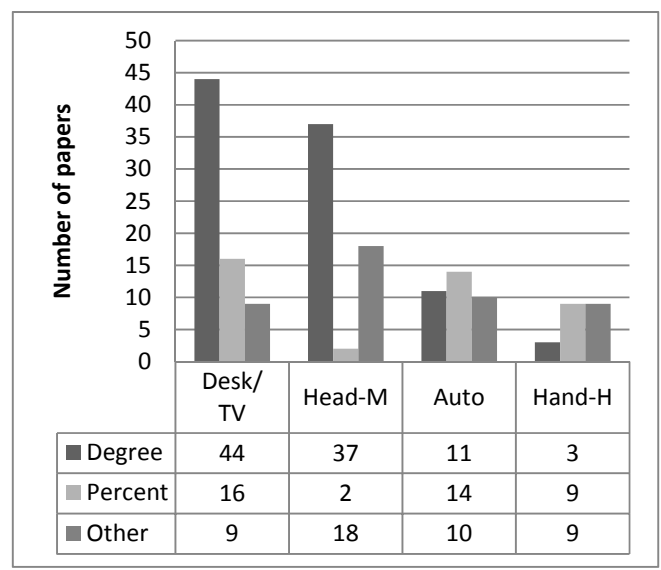

Fig.10. Diversity in metrics used for representing performance in different platforms (Head-M is head-mounted, Auto is for automotive systems and Hand-H represents handheld devices like smartphones and tablets that implement gaze tracking). The figure shows that although degree measures are most common format of measurement, results are reported in several other formats which cannot be linked to each other. The inhomogeneity is majorly observed in automotive and handheld platforms.

TABLE V
DIVERSITY IN PERFORMANCE METRICS IN GAZE ESTIMATION SYSTEMS

| Platform | No. of papers | References where following performance metrics are used | | |
|---|---|---|---|---|
| | | Degree (95 papers) | Percentage (41 papers) | Others (46 papers) |
| **Desktop and TV panels** | 69 | [5][65][67][68][152][6][66] [165][166][41] [44][42][47] [56][48][43][45][57][54][55] [4][53][148][58][147][167] [146][37][91][168][145][79] [150][8][169][51][52][62][63][64] [85][87][71][72] | [76] [83] [82][73][88] [141][93][95][96][170] [171][94][172] [84][75][80] | [74][89][97][173][92] [174][98][153][77] |
| **Head mounted** | 57 | [101][100][9][10][175] [176][177][178][102][106] [104][179][180][165][181] [182][105][107][183][109] [106][103][184][185][12] [154][186][187][156][188] [22][155][21][114][115][109][50] | [189] [117] | [11][176][190][191][192] [193][194][195][196][197] [198][35][199][200][156] [110] [111][112] |
| **Automotive** | 35 | [13][157][123][16][158][201] [202][131][128][129][132] | [203][162][15][17][122] [121][125][119][120][16] [127][160][204][133] | [14][205][206][159][124][161] [207][204][208][130] |
| **Hand-held** | 21 | [136][137][18] | [139][209][164][210][163] [143][142][140][134] | [211][212][141][108][213][135] [19][144][86] |

*B. Platform specific factors affecting usability of gaze tracking systems*

Eye gaze estimation systems on various user platforms and applications face a broad range of operating conditions that are rarely taken into consideration or characterized during their development. The practical performance of a gaze tracking system in reality may be affected by several factors (or we may call them: sources of errors) that are common or unique to each platform. Some of these factors and their effects have been discussed in our earlier work [24].

For example, the major sources of errors in desktop are head movements. In head-mounted trackers, errors may arise due to "Midas-touch" effect[31], miss-calibration or tracker latency. Errors in automotive systems may arise from platform and user head movements, variable illumination, occlusion due to shadows or user wearing glasses. In handheld devices, eye tracking gets highly challenged by changing positions of the user with respect to the device, head pose, hand jitter, variable illumination, and Midas-touch.

In gaze research, effects of some of these conditions such as head-pose changes, user distance and viewing angle are studied but other factors such as display properties i.e. size and pixel resolution of the screen where gaze is tracked, platform movement and jitter, illumination changes, camera quality and human eye limitations are very sparsely reported. To study this scenario, Table VI presents an information statistics on practical conditions that may affect gaze tracking and the extent to which they have been studied. Firstly the different error sources are listed in Column I. Then the papers are categorized according to platforms to identify the research works where these error sources have been reported. This information is presented in the different columns of Table VI.

It is seen from the table that each user platform encounters at least 5-6 different conditions which can affect their performance but their impacts are very sparsely studied for all user platforms. Head-pose is the most studied factor out of all but very few papers analyze the impact of others error sources.

Further in presence of these factors, the real performance of an eye tracker can deviate significantly and unpredictably from the scores reported under ideal conditions. Hence, unless these error sources are adequately evaluated, the accuracy achieved by a gaze tracking system cannot be reliably specified.

VI. A PERFORMANCE EVALUATION FRAMEWORK FOR EYE GAZE ESTIMATION SYSTEMS

*A. Need and rationale for developing comprehensive performance evaluation strategies for gaze estimation systems*

From the sections and tables above it is seen that currently there exists a large diversity among gaze tracking methods, setup, implementation platforms and accuracy metrics.

An important issue that becomes apparent is that there is currently no comprehensive practice for realistic performance evaluation of gaze tracking systems. Most research works do not assess their systems under the impact of various error sources or provide adequate details about their system configurations. For example, Table VI shows that only 35 out of 69 papers on desktop based systems and 16 out of 57 papers on head-mounted systems report effect of head pose variations. Only 2 works in each of desktop and head-mounted platforms report effect of display properties of the gaze tracking setup. Effect of illumination changes is reported in 4 papers in desktop and one in head-mounted systems. In dynamic platforms like automotive and hand-held devices, where external conditions are more variable, the evaluation statistics is even poorer, as seen from Table VI. Another issue observed is that the accuracy scores for gaze based systems are presented using varied formats and metrics, such as angular resolution, correct detection rate, pixel and physical distances etc. which makes them difficult to interpret and inter-compare. For example: Table V and Fig. 10 show that out of 182 total research works across all platforms, 95 papers report gaze accuracy scores in degrees whereas 41 papers report in percentages (correct detection rates) and 46 papers use other heterogeneous units.

TABLE VI
SOURCES OF ERRORS IN GAZE ESTIMATION SYSTEMS AND LITERATURE WHERE THEY ARE REPORTED

| Error Sources | Platforms | | | |
|---|---|---|---|---|
| | Head Mounted | Desktop (including TV) | Automotive | Handheld (smartphones, tablets) |
| Head pose/ movement | [214][104][21][181] [177][22][9][106] [176][186][184][178] [183][35][109] [117] | [5][65][67][68][152] [6][66][42][44][46] [47][56][48][43][45] [57][54][55][4][148] [167][37][91][81][145] [149][74][150][99][173] [62][63][64][87][71] | [203][162][14][15] [126][159][124][13][121][125] [157][119][16][160][133][131] [128][130] | [108][144][86] |
| Camera quality/image resolution | [117] | [65][54][49][55] [53][145][75] | [no ref] | [no ref] |
| Display properties | [194][199] | [54][55] | [129] | [no ref] |
| User distance/ Viewing angle | [192][154][165][109] [103] [117] | [65][67][68][6][47] [43][45][55][4][169] [37][97][51][63] | ---- | [164][163][86] |
| Hand/Platform movement | [114][50] | -- | X [no ref] | X [141][140][86] |
| Illumination changes | [111] | [5][91][79][87] | [162][133][130] | [no ref] |
| Occlusion | [105][102] | [5][67][7][169][71] | [126][127][133][128][130] | [no ref] |
| Human eye conditions/user dependence | [111][115] | [65][152][6][45] [169][145][95][58] | [128][129] | [213] |

There is also considerable ambiguity in terminologies used in eye gaze research. For example, mentioning "slight head movement" (as in [78], [145], [43]) or "large head movement" (in [47], [66], [45]) and free head movement (in[56], [53],[67]) gives no quantitative idea about the real extent of head pose variations tolerated by these gaze tracking systems.

Considering these factors, the development of comprehensive evaluation strategies for gaze tracking systems seems necessary for several reasons: a. to study impact of various error sources on system performance b. to report system performance quantitatively in uniform formats c. compare results from different eye tracking systems under different operating conditions d. to identify the main bottlenecks for each platform. We present here the concept of such an evaluation framework in the sections below.

### B. Concept of a performance evaluation framework for gaze estimation systems

The framework is to be built around a set of standardized experiments for evaluating various gaze tracking systems. Through the experiments, practical performance limits of a given gaze tracking algorithm or system can be tested under the influence of various parameters: such as variations of head pose, viewing angle, screen size and resolution, eye-occlusion, platform movement and illumination changes. The structure of the framework is outlined in Fig.11. The advantage of having such a framework is that it can answer several critical queries related to eye gaze system design and performance.

For example, which system parameters affect the performance in a particular use-case? How does a particular system perform when compared with similar systems under certain operating conditions or in a particular use-case? Can an algorithm designed for one platform be ported and implemented effectively in another platform? What are the performance bottlenecks of individual algorithms? At present it is challenging to answer such questions as there are no resources that allow us to do practical comparative testing of gaze estimation systems.

### C. Methodology

A typical eye tracking setup comprises of the user, gaze tracker and the tracking environment and each of these components may influence the overall eye tracking performance. A schematic diagram of such a setup and these factors are listed in Fig.11. The proposed experimental framework aims to test impacts of these factors on a tracker's accuracy. A typical "experiment" consists of the following steps: a user is asked to sit in front of the eye tracker and their eyes are calibrated for a session. The user is presented with a graphical user interface when the tracker records their gaze coordinates as the user gazes at several points on the screen. The gaze error in degrees is calculated from the shift between ground truth and tracked gaze locations. Some evaluation experiments done and planned with commercial eye trackers are presented here. Preliminary results can be found in [24].

*a) Estimating impact of head pose*

A user is positioned in front of the eye tracker while a video camera captures the position of the user's head simultaneously. The user's head pose in roll, pitch and yaw angles are obtained from the video using an appearance model as shown in Fig.12. Then the user is asked to turn their head to specific fixed positions (in roll pitch yaw angles) and their gaze is tracked on the same interface again. The gaze accuracy scores corresponding to various head pose angles are presented in Fig 13. For a particular tracker it was observed that for reliable gaze tracking, head pose must be restricted within 20 degrees of movement in 3 directions and in this way, the practical head-pose tolerance limits of the tracker could be estimated.

*b) User distance and viewing angle*

For this experiment, the users are positioned at successively increasing distances from the tracker - computer screen setup (40 cm to 100 cm in 15 cm interval) and the gaze tracking accuracy data are recorded for each user position for fixed frontal head poses. The user viewing angle decreases as the user moves away from the screen and the gaze tracking errors are seen to increase with increasing viewing angles. Tracking stops when users are closer than 40 cm.

*c) Illumination levels*

Several different illumination levels can be introduced during the experiments. There are cool temperature fluorescents (color temperature ~6400K), warm incandescent lamps (color temperature ~ 2500 K) and mixed lamps (color temperature ~ 5500K) that are available at various intensities (100-3500 lux). Eye tracking experiments will be repeated under these illumination levels to study impact on tracking accuracy.

*d) Display size and resolution*

The eye tracking experiment is run on displays of various sizes (9, 11, 13.5, 15 inches) and resolutions (800x600, 1024x768, 1280x768, 1366x768) and the corresponding gaze tracking errors are estimated. We have plans for including a 44 inch TV with a specialized tracker intended for gaze tracking for large screens in our testing framework and study impacts of user viewing angles and display sizes.

*e) Occlusion*

We have already observed certain eye trackers having problems with data collection from a user wearing glasses while another tracker has no such issues. Therefore, the level of tolerance of a tracker to user wearing glasses have to be evaluated by operating both trackers simultaneously while doing the gaze experiment for users with and without glasses.

*f) Platform movements*

The operating conditions faced by a gaze tracker on a static platform like desktop is largely different from that of a dynamic platform like a smartphone or head-mounted setup. We therefore plan to evaluate performance of eye trackers running on such dynamic platforms to realistically observe the difference in performance, study impact of platform movements and other influencing factors.

### D. Studying dynamic eye movement characteristics

In reality, during a given task the human eyes are constantly moving and therefore sequential eye movements can be studied as a statistical process. Through our framework, we aim to do dynamic eye measurements in a video for studying smooth pursuits besides regular fixations. This is a planned inclusion in which the user will be presented with a 'moving target' for the eyes to follow while capturing video of the eye-movements.

This framework is under development at the moment and preliminary results from some of the experiments have been published by us in [24]. The focus of this paper is mainly to highlight the issue of realistic performance evaluation of eye gaze systems through the literature review. Therefore only limited details about the methodology and implementation of the framework are presented here. A more comprehensive paper on the technical details of the framework is under preparation and will have more information.

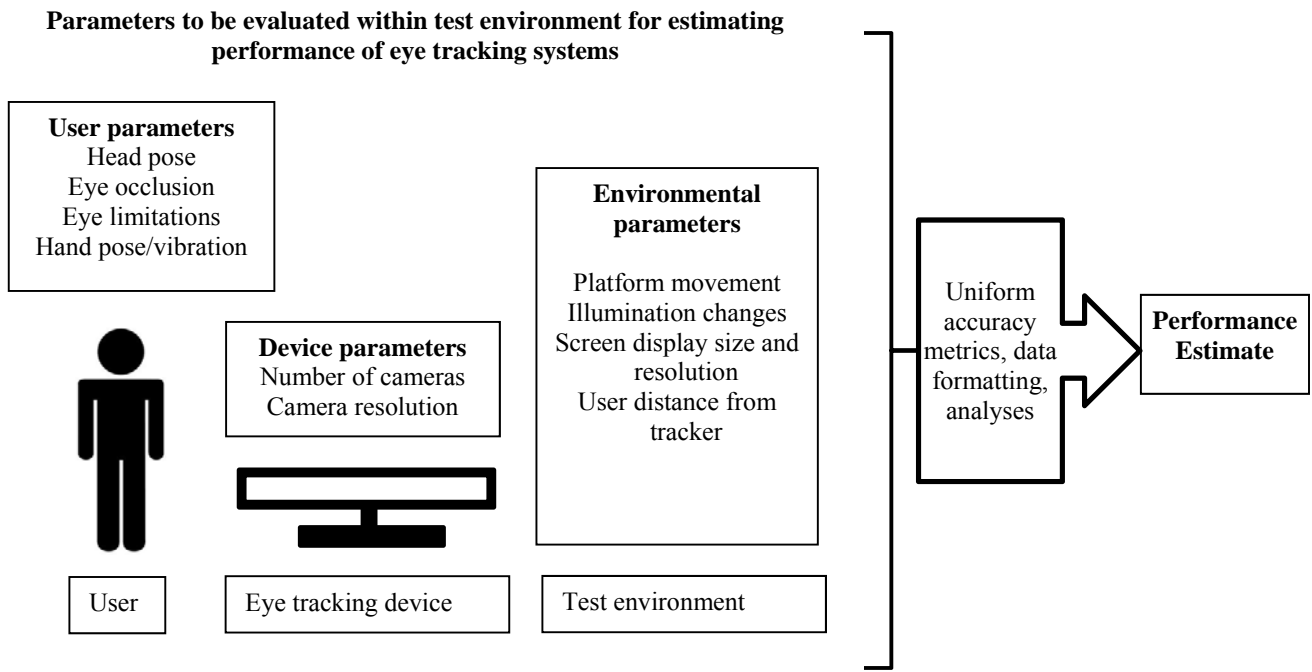

Fig.11. Outline concept of a methodological framework for performance evaluation of eye gaze systems

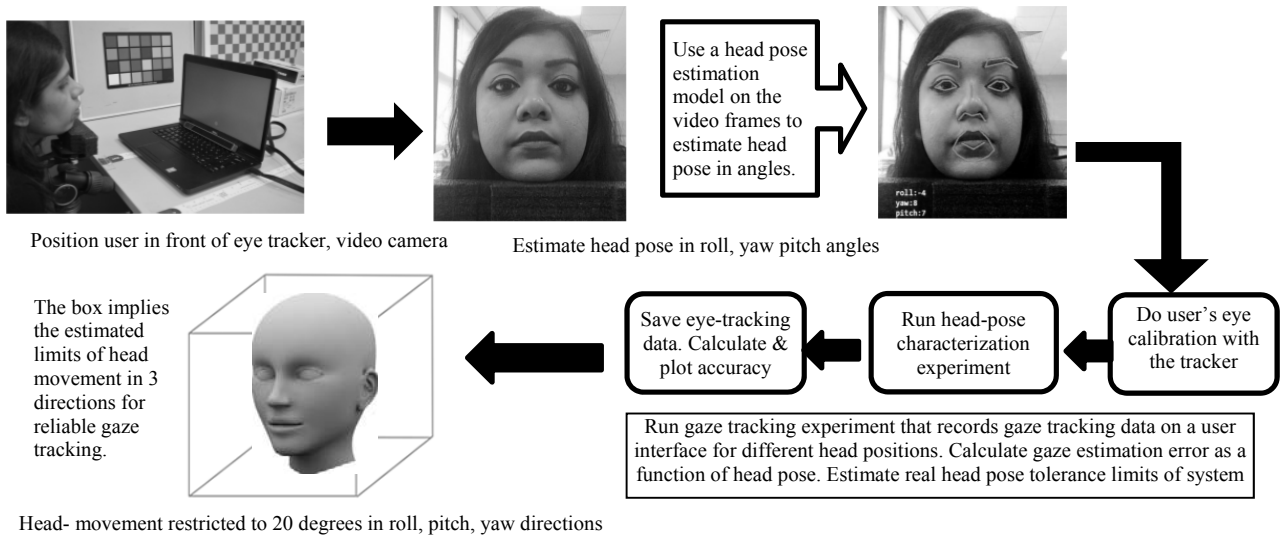

Fig.12. Steps of quantifying head pose tolerance limits of a given eye tracking system

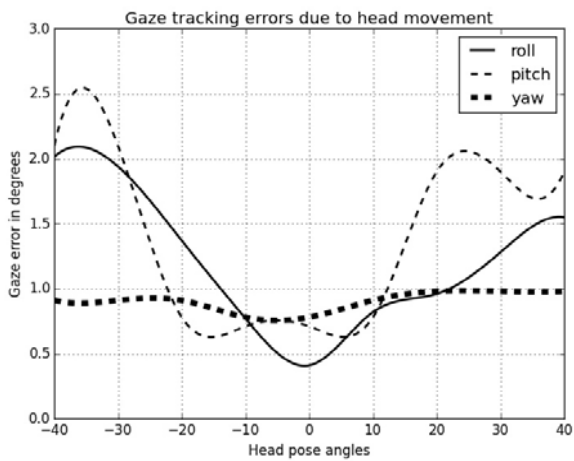

Fig.13. The plot shows the gaze tracking errors arising due to user head motion while using a commercial eye tracker during our experiments. The human head has a degree of flexibility in the order of ~40 degrees of angular movement in each of the roll, pitch and yaw directions. However the plot implies that for reliable gaze tracking with the given tracker, a user needs to keep their head position limited within a fixed angular range in each of these three directions. Through our evaluation process, we estimated this head movement limitation to be equivalent to ~20 degrees about the central position. If the user's head moves beyond these angles, gaze tracking errors of the tracker rises above acceptable levels. However, another feature observed from this plot is that the impact of head roll and pitch on tracking error are relatively more pronounced than yaw variations.

## VII. CONCLUSION AND FUTURE WORK

Eye gaze estimation is an interdisciplinary area of research and development which has received quite a lot of interest from academic, industrial and general user communities in the last decades owing to the ease of availability of computing and hardware resources and increasing demands for human computer interaction methods. In this paper, a detailed literature review is made on the recent advances in eye gaze research, and information in statistical format is presented to highlight the diversity in various aspects such as platforms, setups, users, algorithms and performance measures existing between different branches of this field.

Several different gaze tracking algorithms and their respective advantages and disadvantages were analyzed in Section III. Currently gaze based HCI systems are capable of achieving high speed input and control operations, leading to their implementation in a variety of user platforms and applications, which were discussed in Section IV. Typically, gaze tracking systems at present are capable of determining 3D point of gaze in real time with unconstrained head movement and achieve around 0.5 degrees of angular resolution. However, limitations arising due to gaze tracking camera quality, random illumination changes, user wearing glasses and platform vibrations are not well characterized in contemporary eye gaze research.

The literature review also raises a major question with respect to the consistency and accuracy that can be obtained from the gaze estimation systems when they operate under real world conditions, if they are not properly evaluated.

A variety of factors may affect eye gaze tracking in different platforms, making their performance unpredictable and ultimately questioning their usability in present and future applications. Effects of head movement, user distance and viewing angle, display properties of the setup are still poorly studied, as discussed in Section V. In their presence, practical system performance may differ significantly from expected values and eye gaze may lose its applicability in different consumer use cases.

Further, there is a clear lack of homogeneity in gaze performance metrics as pointed out in the tables of Section V. Some performance measures used, for example: detection rate or accuracy percentage is difficult to interpret physically and the variety in reporting formats makes inter-comparisons between different systems and algorithms impossible.

Keeping these in mind, the concept of a performance evaluation framework is proposed that will provide practical performance estimates of gaze tracking systems and adopt a uniform set of accuracy metrics for specifying performance. This is an on-going research activity and the details of the evaluation methods to be included in this framework are currently under development for different gaze estimation platforms, and will be included in a subsequent paper.

ACKNOWLEDGMENT

The authors would like to thank Prof. Christopher Dainty and Dr. Claudia Costache for reviewing and providing their valuable feedback on the manuscript.

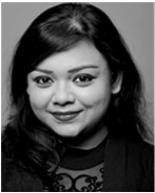

**Anuradha Kar** is currently a PhD student at the National University of Ireland Galway and is with the Center for Cognitive, Connected, and Computational Imaging (http://www.c3imaging.org). Her research interests include human computer interaction (HCI) and computational imaging. Currently she is working in the area of eye gaze tracking - addressing the issues of accuracy and performance evaluation of gaze estimation systems in various platforms.

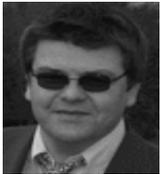

**Peter Corcoran** (F'10) has co-authored over 300 technical publications and co-inventor on more than 250 granted US patents. His research interests include biometrics, cryptography, computational imaging, and consumer electronics. He is the Editor-in-Chief of the *IEEE Consumer Electronics Magazine* and a Professor with a Personal Chair at the College of Engineering & Informatics at NUI Galway. In addition to his academic career, he is also an Occasional Entrepreneur, Industry Consultant, and Compulsive Inventor.